\documentclass[english,twocolumn,prl,showpacs,preprintnumbers,amsmath,amssymb]{revtex4}
\usepackage[T1]{fontenc}
\usepackage[latin1]{inputenc}
\usepackage{amsmath}
\usepackage{graphicx}

\makeatletter
\@ifundefined{textcolor}{}
{%
 \definecolor{BLACK}{gray}{0}
 \definecolor{WHITE}{gray}{1}
 \definecolor{RED}{rgb}{1,0,0}
 \definecolor{GREEN}{rgb}{0,1,0}
 \definecolor{BLUE}{rgb}{0,0,1}
 \definecolor{CYAN}{cmyk}{1,0,0,0}
 \definecolor{MAGENTA}{cmyk}{0,1,0,0}
 \definecolor{YELLOW}{cmyk}{0,0,1,0}
 }

\makeatletter
\usepackage{epsfig}

\usepackage{dcolumn}

\usepackage{bm}

\usepackage{color}

\makeatletter

\usepackage{ulem}

\makeatother

\makeatother

\makeatother

\usepackage{babel}

\begin{document}

\title{A new type of vacancy-induced localized states in multilayer graphene}

\author{Eduardo V. Castro,$^{1,2}$ María P. López-Sancho,$^{1}$ María A.
H. Vozmediano$^{1}$}

\affiliation{$^{1}$Instituto de Ciencia de Materiales de Madrid, CSIC, Cantoblanco,
E-28049 Madrid, Spain}

\affiliation{$^{2}$Centro de Física do Porto, Rua do Campo Alegre 687, P-4169-007
Porto, Portugal}
\begin{abstract}
We demonstrate the existence of a new type of zero energy state associated
to vacancies in multilayer graphene that has a finite amplitude over
the layer with a vacancy and adjacent layers, and the peculiarity
of being quasi-localized in the former and totally delocalized in
the adjacent ones. In a bilayer, when a gap is induced in the system
by applying a perpendicular electric field, these states become truly
localized with a normalizable wavefunction. A transition from a localized
to an extended state can be tuned by the external gate for experimentally
accessible values of parameters. 
\end{abstract}


\maketitle
%-----------------------------------------------------------------------------%
%-----------------------------------------------------------------------------%

Graphene is a one atom thick layer of carbon atoms ordered in a honeycomb
lattice. The enormous interest risen since its discovery \citep{NGM+04}
is driven equally by potential technological applications \citep{PSK+08}
and unconventional low-energy behavior (massless Dirac quasi-particles)
\citep{NGPrmp}. Together with single layer graphene (SLG), bilayer
graphene (BLG) and multilayer graphene (MLG) structures were also
synthesized. The BLG structure being unique as a low energy effective
model \citep{MF06}, rises better technological expectations due to
the possibility to open and tune a gap in the spectrum by electric
field effect \citep{McC06,CNM+06,OHL+07}. A proper understanding
of the effect of disorder is crucial for technology relevant applications.
Annealing and removing the substrate has recently lead to an increase
in mobility by one order of magnitude in SLG \citep{BSJ+08,DSB+08}.
Intrinsic defects, such as vacancies or topological lattice defects
are not easy to get rid of and further investigation on their role
is mandatory. Vacancies are lately been recognized as one of the most
important scattering centers in SLG and BLG \citep{MAW+09,GBetal08}
and the zero modes induced by this type of defects can greatly affect
the transport properties of the samples as well as the possible electronic
instabilities near the neutrality point.

\begin{figure}[t]
\begin{centering}
\includegraphics[width=0.99\columnwidth]{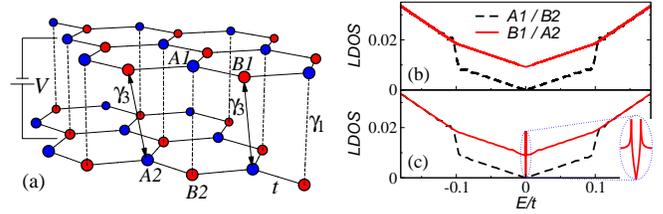} 
\par\end{centering}

\caption{\label{fig:bilLattLDOS}(color online). (a)~Bilayer lattice structure
and main tight-binding parameters. (b)-(c)~LDOS for $\gamma_{3}=0$
and $\gamma_{3}=0.1t$, respectively.}

\end{figure}

In the present manuscript we address the character of vacancy-induced
electronic states in BLG, with extension to the case. We begin by
summarizing the main findings of this work. (i) For the minimal model
we construct an analytic solution on the lattice for the zero modes
associated to the two different types of vacancies in BLG -- located
at $A1/B2$ or $B1/A2$ {[}see Fig.~\ref{fig:bilLattLDOS}(a){]}.
We demonstrate that a new type of state different from these found
in SLG and in other layered systems exists in BLG. The peculiarity
consists in having a finite amplitude over the two layers and, more
exotic, the wave function is quasi-localized in one layer and totally
delocalized in the other. We also prove that these states survive
in the continuum limit. The states associated to the $A1/B2$-vacancies
are quasi-localized decaying as 1/r at large distances, similar to
these of SLG, while those corresponding to a $B1/A2$-vacancy are
delocalized. The delocalization is due to the spread of the wave function
in the opposite layer where the vacancy resides. This solution is
directly applicable to MLG and graphite. (ii) We demonstrate that
these localization properties survive in the presence of non-minimal
coupling $\gamma_{3}$ by means of a numerical analysis including
the study of the participation ratios. (iii) We study the behavior
of these states in the presence of a gap and find that these associated
to the $B1/A2$-vacancy become truly localized states leaving inside
the gap while the $A1/B2$-vacancy (monolayer type) become delocalized.
The truly localized states are located symmetrically around the middle
of the gap -- depending on the layer they belong to.

\emph{Model.}---The tight-binding minimal model for the $\pi-$electrons
in $AB$-stacked BLG is schematically shown in Fig.~\ref{fig:bilLattLDOS}(a).
We will use the parameters $t\approx3\,\mbox{eV}$ and $\gamma_{1}\approx t/10$
\citep{NGPrmp}. It has two parabolic bands that touch at two degenerate
Fermi points with a constant density of states (DOS) at the Fermi
points. For the present study of zero modes it will also be of interest
to consider the interlayer hopping $\gamma_{3}\approx\gamma_{1}/3$
that linearizes the bands around the Fermi points and induces a vanishing
DOS at zero energy. Figure~\ref{fig:bilLattLDOS} shows the local
DOS (LDOS) for $\gamma_{3}=0$~(b) and $\gamma_{3}\neq0$~(c). The
presence of a finite gap induced through a perpendicular electric
field $E_{z}=V/(ed)$, where $d\approx0.34\,\mbox{nm}$ is the interlayer
distance, is included by adding an on-site energy term: $-V/2$ at
layer~1 and $V/2$ at layer~2. Within the present model vacancies
correspond to the elimination of lattice sites. We do not include
any reconstruction of the remaining structure. Even though some reconstruction
might be present in real systems, the zero-energy modes we are interested
in here seem to be rather insensitive to it \citep{CJK+08}.

%-----------------------------------------------------------------------------%
%-----------------------------------------------------------------------------%

%
\begin{figure}[t]

\begin{centering}
\includegraphics[width=0.98\columnwidth]{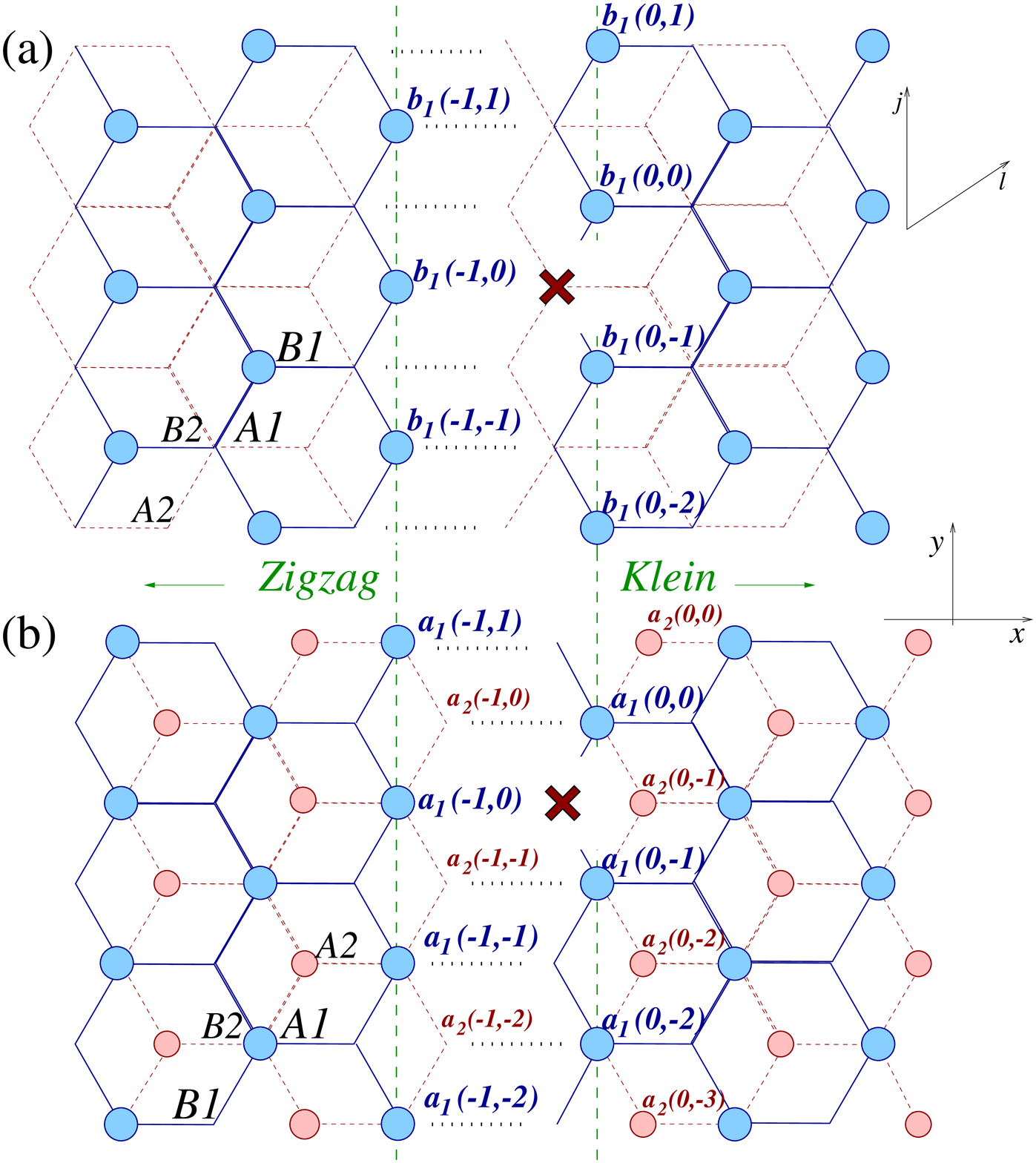} 
\par\end{centering}

\caption{\label{fig:zigzagKlein}(color online). Schematics for constructing
a vacancy-induced zero-energy solution in bilayer graphene (see text).
Circles indicate sites where the localized states have a finite amplitude.
(a)~$A1/B2$ vacancy. (b)~$B1/A2$ vacancy.}

\end{figure}

\emph{Analytic construction of the vacancy states.---}A vacancy in
the honeycomb lattice gives rise to a quasi-localized state \citep{PGS+06}
whose wave function can be written as \begin{equation}
\Psi(x,y)\approx\frac{e^{i\mathbf{K}.\mathbf{r}}}{x+iy}+\frac{e^{i\mathbf{K}'.\mathbf{r}}}{x-iy},\label{eq:PsiVac1L}\end{equation}
where $\mathbf{K}$ and $\mathbf{K}'$ are the reciprocal space vectors
of the two inequivalent corners of the first Brillouin zone, and $(x,y)$
are distances in a reference frame centered at the vacancy position.
We will construct an analytic solution for vacancy-states in BLG within
the minimal model following the analysis done for SLG in \citep{PGS+06}.
The wave function is obtained by matching surface state solutions
at zigzag edges with those localized at Klein edges for a suitable
boundary condition. The schematics used in this construction is shown
in Fig.~\ref{fig:zigzagKlein}. The amplitude of the zero-mode wavefunction
is denoted $c_{i}(l,j)$, with $c=a,b$ and $i=1,2$ for sites in
sublattice $A,B$ and layer $1,2$ of the unit cell located at $(l,j)$.
By cutting the system into \emph{left} and \emph{right} regions, defined
with respect to the vacancy position, we see that in order to have
a solution that decays away from the vacancy we need a zigzag-edge
surface state to the left and a Klein-edge surface state to the right.
The existence of surface states localized at zigzag edges in BLG has
been proven in Ref.~\citep{CPL+07}. There are two linearly independent
solutions, one living in a single layer (monolayer type) and the other
having a finite amplitude over the two layers (bilayer type). Only
sites belonging to the sublattice containing the zigzag edge have
a finite amplitude. In an analogous way, one can show that surface
states localized at Klein edges in BLG exist as well \citep{foot1}.
Again, two linearly independent solutions show up (monolayer and bilayer
types).

Consider a vacancy at $A1/B2$ sites, as sketched in Fig~\ref{fig:zigzagKlein}(a).
The zigzag- and Klein-edge surface states to be used have to have
a finite amplitude, respectively, on the first zigzag column to the
left and on the first beard column to the right of the vacancy. This
is imposed by the matching (boundary) condition, which reads\begin{equation}
b_{1}(-1,j)+b_{1}(0,j)+b_{1}(0,j-1)=0\label{eq:bc}\end{equation}
 for all $j$'s except at the vacancy, and involves sites of the two
mentioned columns. The zigzag- and Klein-edge states with amplitudes
starting at these columns are those of the monolayer type, i.e. with
weight only on one layer:\begin{eqnarray}
b_{1}(l<0,j) & = & \sum_{k_{m}}b_{1}(-1,k_{m})D_{k_{m}}^{-(l+1)}e^{ik_{m}(\frac{l+1}{2}+j)},\label{eq:ZZ}\\
b_{1}(l\geq0,j) & = & \sum_{k_{m'}}b_{1}(0,k_{m'})D_{k_{m'}}^{-l}e^{ik_{m'}(\frac{l}{2}+j)},\label{eq:KL}\end{eqnarray}
 where $D_{k}=-2\cos(k/2)$, and the sums go over $2\pi/3\leq k_{m}\leq4\pi/3$
in Eq.~(\ref{eq:ZZ}) and $0\leq k_{m'}\leq2\pi/3$ and $4\pi/3\leq k_{m'}\leq2\pi$
in Eq.~\eqref{eq:KL}, for momenta $k_{m},k_{m'}$ along the $y-$direction,
with $b_{1}(l,k)$ the Fourier transform of $b_{1}(l,j).$ The analysis
now is completely analogous to the SLG case~\citep{PGS+06}. Namely,
the boundary condition~(\ref{eq:bc}), conveniently rewritten as
$\sum_{k_{m}}b_{1}(-1,k_{m})e^{ik_{m}j}=-\sum_{k_{m'}}(1+e^{ik_{m'}})b_{1}(0,k_{m'})e^{ik_{m'}j}$,
is satisfied for all $k_{m}$ and $k_{m'}$ in the ranges indicated
above by choosing $b_{1}(-1,k_{m})=1$ and $b_{1}(0,k_{m'})(1+e^{ik_{m'}})=1$.
Going from lattice indices $(l,j)$ to distances $(x,y)$ we obtain
exactly the result given by Eq.~\eqref{eq:PsiVac1L}. Therefore,
for a vacancy at $A1/B2$ sites in BLG a quasi-localized (decaying
as $1/r$) zero-energy mode exists around the vacancy, living in the
same layer but opposite sublattice.

Consider now a vacancy at $B1/A2$ sites, sketched in Fig.~\ref{fig:zigzagKlein}(b).
The zigzag- and Klein-edge states with a finite amplitude, respectively,
over sites $(-1,j)$ and $(0,j)$ of layer~1, are now those of the
bilayer type. These states have amplitudes over layer~1 still given
by Eqs.~(\ref{eq:ZZ}) and~(\ref{eq:KL}), with the replacement
$b\rightarrow a$. Additionally, they have also finite amplitudes
over layer~2, which can be written as \begin{eqnarray}
a_{2}(l<0,j) & = & \frac{\gamma_{1}}{t}\sum_{k_{m}}a_{1}(-1,k_{m})(l+1)D_{k_{m}}^{-(l+2)}e^{ik_{m}(\frac{l+2}{2}+j)},\label{eq:ZZ2}\\
a_{2}(l\geq0,j) & = & \frac{\gamma_{1}}{t}\sum_{k_{m'}}a_{1}(0,k)(l+1)D_{k_{m'}}^{-(l+1)}e^{ik_{m'}(\frac{l+1}{2}+j)},\label{eq:KL2}\end{eqnarray}
 with momenta $k_{m},k_{m'}$ restricted to the intervals mentioned
before. An important point to note is that the boundary condition
reads exactly the same as in Eq.~(\ref{eq:bc}), with the replacement
$b\rightarrow a$. Even though we are using zigzag- and Klein-edge
states which have finite amplitudes in both layers, it happens that,
by construction, the weight~(\ref{eq:ZZ2}) of the zigzag surface
state at layer~2 is such that $a_{2}(-1,j)=0$, and thus the matching
condition at this layer is satisfied by default. At this point the
derivation follows closely that for a vacancy at $A1/B2$. Noting
that in layer~1 we have to match exactly the same edge-state solutions
given by Eqs.~(\ref{eq:ZZ}) and~(\ref{eq:KL}), with $b\rightarrow a$,
and that in layer~2 Eqs.~(\ref{eq:ZZ2}) and~(\ref{eq:KL2}) can
also be written in the same form as Eqs.~(\ref{eq:ZZ}) and~(\ref{eq:KL}),
apart from the term $(l+1)\gamma_{1}/t$, we arrive at the following
zero-mode behavior, \begin{equation}
\Upsilon(x,y)\sim\Psi(x,y)\left[1,x\,\gamma_{1}/t\right],\label{eq:Upsilon2L}\end{equation}
 where $\Psi(x,y)$ is the quasi-localized state given in Eq.~(\ref{eq:PsiVac1L}),
and the two component wave function refers to the two layers; first
and second components for the first and second layers, respectively.
This is a delocalized state, with the peculiarity of being quasi-localized
in one layer (where the vacancy sits) and delocalized in the other
where it goes to a constant when $r\rightarrow\infty$.

The analytic construction used for the minimal model in BLG applies
directly to MLG and graphite with Bernal stacking along the lines
of Ref.~\citep{CPS08}. The quasi-localized state~\eqref{eq:PsiVac1L}
is a solution in any multilayer with a $A1/B2$-vacancy. For a $B1/A2$-vacancy
the solution is a generalization of state~\eqref{eq:Upsilon2L} with
a quasi-localized component in the layer where the vacancy resides
and delocalized components in the layers right on top and below this
one: $\Phi(x,y)\sim\Psi(x,y)\left[1,x\,\gamma_{1}/t,x\,\gamma_{1}/t\right]$.

\emph{The continuum limit.---} Both the conventional {[}Eq.~\eqref{eq:PsiVac1L}{]}
and the unconventional {[}Eq.~\eqref{eq:Upsilon2L}{]} solutions
are fully consistent with the low-energy approximation for BLG \citep{MF06}.
Far from the vacancy the zero modes must obey $\partial_{\bar{z}}^{2}\psi_{B1}(z,\bar{z})=0$
and $\partial_{z}^{2}\psi_{A2}(z,\bar{z})=0$ at $\mathbf{K}$, where
$z=x+iy$ and $\bar{z}=x-iy$, and a similar set at $\mathbf{K}'$
with $z$ replaced by $\bar{z}$ everywhere. An obvious solution has
$\psi_{B1}(z,\bar{z})=f(z)$ and $\psi_{A2}(z,\bar{z})=0$, or $\psi_{B1}(z,\bar{z})=0$
and $\psi_{A2}(z,\bar{z})=f(\bar{z})$, with $f(z)$ analytic. Adding
the contribution of the two $K$'s we see that Eq.~\eqref{eq:PsiVac1L}
is precisely of this form; the amplitude over the sublattice opposite
to the vacancy behaving as $1/z+1/\bar{z}$, analogous to the quasi-localized
solution in SLG \citep{PGS+06}. Interestingly, the bilayer model
also supports solutions with $\psi_{B1}(z,\bar{z})=\bar{z}f(z)$ and
$\psi_{A2}(z,\bar{z})=0$, or $\psi_{B1}(z,\bar{z})=0$ and $\psi_{A2}(z,\bar{z})=zf(\bar{z})$.
Equation~\eqref{eq:Upsilon2L} at the low-energy sublattice opposite
to the vacancy is indeed a combination of the stated solutions, namely
$\bar{z}/z+z/\bar{z}$ \citep{foot3}.

%-----------------------------------------------------------------------------%
%-----------------------------------------------------------------------------%

%
\begin{figure}[t]

\begin{centering}
\includegraphics[width=0.9\columnwidth]{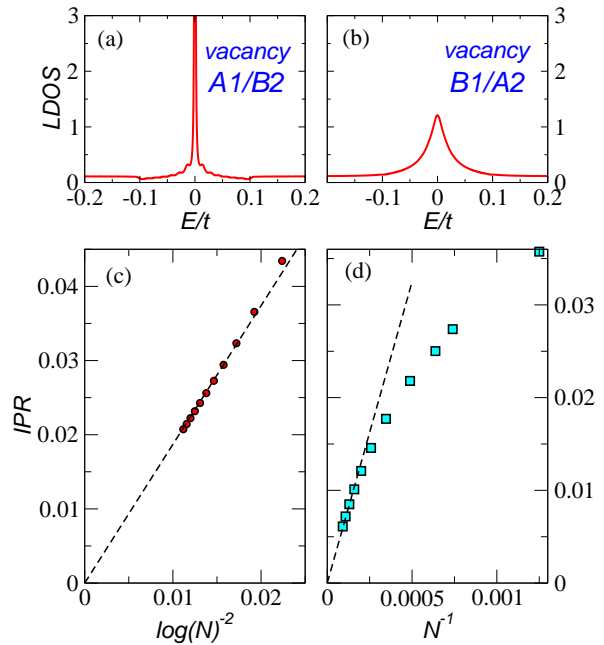} 
\par\end{centering}

\caption{\label{fig3}(color online).  LDOS~(a)-(b) and IPR~(c)-(d) for a
vacancy at sublattice $A1/B2$ (left panels) and $B1/A2$ (right panels).
The LDOS is computed at a lattice site closest to the vacancy. The
IPR is for the zero-energy mode induced by the vacancy. Lines are
guides to the eyes.}

\end{figure}

\emph{Vacancies in the gapless case.---}The analytic results just
presented are for $\gamma_{3}=0$. A finite $\gamma_{3}$ is crucial
for the existence of the quasi-localized state~\eqref{eq:PsiVac1L};
otherwise a finite density of delocalized states exists in the same
energy region {[}see Fig.~\ref{fig:bilLattLDOS}(b)-(c){]}. This
is addressed numerically in the following. We also show that the delocalized
character of the new solution~(\ref{eq:Upsilon2L}) persists in the
presence of a finite $\gamma_{3}$. The localization character of
vacancy-induced modes is studied through finite-size-scaling of the
inverse participation ratio (IPR). The later is defined as $\mathcal{P}_{\nu}=\sum_{i}^{N}|\varphi_{\nu}(i)|^{4}$
for the eigenstate $\nu$, where $\varphi_{\nu}(i)$ is its amplitude
at site $i$. We perform exact diagonalization on small clusters with
$N$ up to $2\times100^{2}$ sites. The IPR for \emph{extended}, \emph{quasi-localized},
and truly \emph{localized} states scales distinctively with $N$ \citep{PdSN07}.
While for extended states we have $\mathcal{P}_{\nu}\sim N^{-1}$,
for quasi-localized states the $1/r$ decay implies $\mathcal{P}_{\nu}\sim\log(N)^{-2}$
(consequence of the definition of the IPR in terms of normalized eigenstates).
For localized wavefunctions the significant contribution to $\mathcal{P}_{\nu}$
comes from the sites in which they lie, and a size independent $\mathcal{P}_{\nu}$
shows up. Additionally to the IPR, we analyze the changes induced
in the LDOS for sites around the vacancy. The LDOS is computed using
the recursive Green's function method in clusters with $N=2\times1400^{2}$,
from which the thermodynamic limit can be inferred.

In Fig.~\ref{fig3}(a) and~\ref{fig3}(b) we show the LDOS at a
lattice site closest to a vacancy located in sublattice $A1/B2$ and
$B1/A2$, respectively. The sharp resonance at zero energy in the
former case is in agreement with the presence of a quasi-localized
state, while the broader feature in the later may be attributed to
the delocalized wavefunction induced by a vacancy in $B1/A2$, which
still presents a quasi-localized component in the layer where the
vacancy sits (and thus the feature). This interpretation is fully
corroborated by the IPR scaling analysis shown in Fig.~\ref{fig3}(c)
and~\ref{fig3}(d) for a vacancy in $A1/B2$ and $B1/A2$, respectively:
quasi-localized state in the former case, and delocalized in the later.

%-----------------------------------------------------------------------------%
%-----------------------------------------------------------------------------%

%
\begin{figure}[t]

\begin{centering}
\includegraphics[width=0.9\columnwidth]{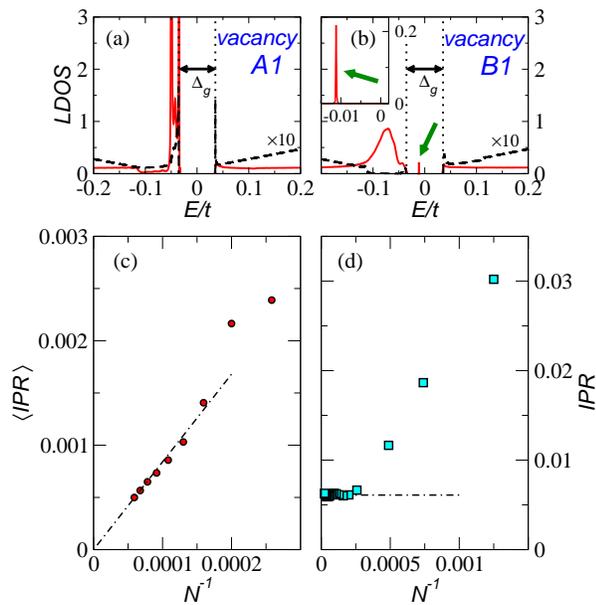} 
\par\end{centering}

\caption{\label{fig:lodsIPRV}(color online). LDOS for a vacancy at sublattice
$A1$ (a) and $B1$ (b) for a finite gap, $V=0.1t$. The LDOS is computed
at a lattice site closest to the vacancy. Dashed lines are for the
perfect lattice. (c) IPR averaged over the gap-edge resonance shown
in (a). (d) IPR for the in-gap mode shown in (b). Dashed-dotted lines
are guides to the eyes.}

\end{figure}

\emph{Vacancies in the gaped case.---}When a finite electric field
$E_{z}$ is present, a gap $\Delta_{g}=[V^{2}\gamma_{1}^{2}/(V^{2}+\gamma_{1}^{2})]^{1/2}$
opens between conduction and valence bands \citep{McC06}. The quasi-localized
state due to a vacancy at sublattice $A1/B2$ becomes a resonance
around $\pm V/2$ in the gaped case, as seen in the LDOS shown in
Fig.~\ref{fig:lodsIPRV}(a) for a site closest to the vacancy. A
strong resonance is seen around $-V/2$ for a vacancy at $A1$ (we
used $V=0.1t$), apart from the known gap edge divergence at $-\Delta_{g}/2$
characteristic of the perfect lattice (dashed line) \citep{GNP06}.
A vacancy at $B2$ gives identical results with $E\rightarrow-E$.
Such a vacancy-induced state living in the continuum is expected to
be delocalized. This is confirmed by the IPR scaling as $N^{-1}$
\citep{foot2}, as shown in Fig.~\ref{fig:lodsIPRV}(c).

For a vacancy at sublattice $B1/A2$, which originates the atypical
delocalized state discussed above when no gap is present, a truly
localized state inside the gap is induced when $E_{z}\neq0$. This
is suggested by the sharp feature seen inside the gap in the LDOS
for a site closest to the vacancy, as shown in Fig.~\ref{fig:lodsIPRV}(b)
and zoomed in the inset (marked by arrows). The IPR scaling to a constant,
as seen in Fig.~\ref{fig:lodsIPRV}(d), fully confirms the localized
nature of this vacancy-induced state. Its asymmetric weight over the
two layers explains why it appears off zero-energy: being negative
for a $B1$ vacancy (layer~1 at an electrostatic energy $-V/2$),
as shown in Fig.~\ref{fig:lodsIPRV}(b), and positive, symmetrically
placed with respect to the center of the gap, for a $A2$ vacancy
(layer~2 at an electrostatic energy $+V/2$).

\emph{Conclusions.---}We have found a new type of zero mode state
in BLG with special features: in the absence of a gap it is quasi-localized
in one of the layers and delocalized in the other and in the presence
of a gap becomes fully localized inside the gap. The results obtained
in this work are directly applicable to MLG and graphite with Bernal
stacking. The findings here reported can be important to understand
recent experiments done in thin films of graphite irradiated with
protons whose main effect is to produce single vacancies on the sample
\citep{ASE+09}. These samples show an enhanced local ferromagnetism
that can be due to the local moments associated to the zero modes
described in this work, and also a better conductivity than the untreated
samples with less defects pointing to the idea that the delocalized
states induced by the vacancies contribute to the conductivity. An
enhanced conductivity has also been found in acid-treated few-layer
graphene \citep{JCW+09}. The localized state found in the gaped case
can also provide a natural explanation for the observation of localization
inside the gap in the biased BLG \citep{OHL+07} and is in agreement
with previous results obtained with impurity models in the continuum
\citep{NN06}.

%-----------------------------------------------------------------------------%
%-----------------------------------------------------------------------------%

We thank F. Guinea, A. Cortijo, and J. M. B. Lopes dos Santos for
useful conversations. This research was partially supported by the
Spanish MECD grant FIS2005-05478-C02-01 and FIS2008-00124. EVC acknowledges
financial support from the Juan de la Cierva Program (MCI, Spain).

%%%%%%%%%%%%%%%%%%%%%%%%%%%%%%%%%%%%%%%%%%%%%%%%%%%%%%%%%%%%%%%%%%%%%%%%%%%%%%%
\bibliographystyle{apsrev}

\begin{thebibliography}{33}
\bibitem[{foo()}]{foot1} \bibinfo{note}{Its analytic form is easily obtained by inverting the $2\times2$   matrix appearing in equation~(4) of reference~\cite{CPL+07}, and following   the steps outlined there.}
\bibitem[{foo()}]{foot3} \bibinfo{note}{Equation~\eqref{eq:Upsilon2L} is a linear combination of  $e^{i\mathbf{K}\cdot\mathbf{r}}\bar{z}/z$ at $K$ and $e^{i\mathbf{K}'\cdot\mathbf{r}}z/\bar{z}$ at $K'$ with the plane waves $e^{i\mathbf{K}\cdot\mathbf{r}}$ and $e^{i\mathbf{K}'\cdot\mathbf{r}}$ -- the two later reflecting the particular zigzag line used to match zigzag and Klein edge states.}
\bibitem[{foo({\natexlab{b}})}]{foot2} \bibinfo{note}{As usually done in regions of continuum DOS, the IPR is an   average over states in an energy bin $\Delta E = 0.2t$ around the resonance.}
\end{thebibliography}
 
\end{document}